\documentclass[11pt,a4paper]{article}
\usepackage{jcappub}
\usepackage{subfigure}
\usepackage{overpic}

\title{Causal structure of accelerating black holes}

\author{Jose A. R. Cembranos,}
\author{Luis J. Garay,} 
\author{Sergio A. Ortega}
\affiliation{Departamento de F\'\i sica Te\'orica and IPARCOS, 
Universidad Complutense de Madrid, 28040 Madrid, Spain}  

\emailAdd{cembra@ucm.es}
\emailAdd{luisj.garay@ucm.es}
\emailAdd{sergioan@ucm.es}

\abstract
{Accelerating black holes are described by the so-called C-metric. In this work, we analyse the causal structure of such black holes by using null geodesics. We construct explicitly the relevant Penrose diagrams. First, we recover well-known results associated with the sub-accelerating
black holes. Then, we extend the study to the super-accelerating case, in which an additional essential singularity appears. In addition, we consider accelerating black holes with negative masses. We show that they are equivalent to the geometry described by the black hole metric beyond conformal infinity.
We compare our results with the Schwarzschild geometry to facilitate understanding and to highlight the interest of the new features.}

\arxivnumber{}

\begin{document}
\maketitle

\section{Introduction}

Black holes are exotic objects described by Einstein's theory of general relativity \cite{Einstein,Einstein2}. They have been a subject of study in recent decades because there exist astrophysical objects with that sort of characteristics, and due to their interesting properties associated with thermodynamics \cite {4laws, Bekenstein} or Hawking radiation \cite {Hawking_1, Hawking_2}.

Accelerating black holes are described by the C-metric, which is a vacuum solution to the Einstein equations discovered by Levi-Civita in 1918 \cite{Levi-Civita}. However, it was not until 1970 that Kinnersley and Walker interpreted the C-metric to describe a pair of black holes uniformly accelerating in opposite directions \cite{Kinnersley}. Subsequently, Plebánski and Demiánski obtained a generalized C-metric with charge and rotation that includes the Kerr-Newman metric \cite{Plebanski}. In recent years, various studies have been carried out with this type of geometries, for example in the study of its thermodynamics \cite{Appels1, Appels2}. 

In this paper we will focus on the causal structure of the spacetime described by the C-metric. This is essential in order to study physical processes in a spacetime. The causal structure has already been studied before in \cite{Diagrams}, where the authors present the Penrose diagrams \cite{Penrose}, but they only consider   accelerations with an upper bound $1/(2m)$ (where $m$ is the black hole mass) which ensure that the acceleration horizon lies outside the event horizon.

We aim to generalize the causal structure for any acceleration value. We will consider  three different situations: sub-accelerating black holes, where the acceleration is less than the upper bound $1/(2m)$; extremal accelerating black hole, where the acceleration equals it; and super-accelerating black holes, where the acceleration surpasses  $1/(2m)$. We will review the construction of the Penrose diagram of the sub-accelerating case, for completeness and because it will help us to understand the analysis in the other cases. We also discuss how the causal structure changes in these three regimes. Finally, we shall also let the mass parameter take a negative value, resulting in a drastically different causal structure of the spacetime.

This paper is structured as follows. Section \ref{sec-c-metric} introduces the C-metric and its main characteristics. In Section \ref{sec-causal} we obtain the Penrose diagrams for this spacetime and for any value of its defining parameters. Finally, we summarize and conclude in Section \ref{sec-conlusions}.

\section{The C-metric}
\label{sec-c-metric}
We are interested in static neutral accelerating black holes in vacuum. It is possible to express the C-metric in pseudospherical coordinates similar to those of the Schwarzschild metric, so they are centred in one of the two accelerating black holes \cite{Griffiths1,Diagrams}. The metric is described by:
\begin{eqnarray}\label{pseudo}
	ds^2&=&\frac{1}{\Omega^2(r,\theta)}\left\lbrace-f(r)dt^2 + \frac{dr^2}{f(r)} +\right.\left.r^2\left(\frac{d\theta^2}{g(\theta)} + g(\theta)\sin^2(\theta)\frac{d\varphi^2}{K^2}\right)\right\rbrace,
\end{eqnarray}where the functions $f$, $\Omega$, and $g$ are given by
\begin{align}
	f(r) &= (1 - A^2r^2)\left(1-{2m}/{r} \right),\qquad
	\Omega(r,\theta) = 1 + Ar\cos\theta,\qquad
%	\label{g}
	g(\theta) = 1 + 2mA\cos\theta,
\label{eq-f}\end{align}
$ m $ is a parameter related to the mass, $ A $ is the acceleration of the black holes,  $\varphi \in [0,2\pi)$ is an angle, and the constant $K$ is related to the conical deficit. We can also take $t \in (-\infty,\infty)$ as the coefficients of the metric do not depend on this variable.

The metric \eqref{pseudo} has  a singularity when the function $g (\theta) = 0$. If $2mA>1$, then $g(\theta)$ on Equation \eqref{eq-f} vanishes for a given value of $\theta$. For this reason, in the literature there is the usual restriction $2mA <1$, so that  $\theta \in [0,\pi)$ \cite{Diagrams}. This is the aforementioned sub-accelerating case. However, as we will see later, when $2mA \geq 1$, then $\theta \in [0,\theta_0)$ with $g(\theta_0) = 0$.

The C-metric presents (in case of $2mA < 1$) conical singularities at the north pole $ \theta_ + = 0 $, and the south pole $ \theta_- = \pi $  associated with the presence of a cosmic string with   tension $   \delta / 8 \pi $, where $ \delta = 2\pi [1 - g(\theta_\pm)/K] $ is the conical deficit \cite{Vilenkin1, String1, String2}. In the literature, it is common to use $K = 1 + 2mA$, so that the conical singularity at the north pole is removed, and a conical deficit remains only at the south pole. This is done because at the north pole the tension of the string, which is proportional to the energy density, would be negative, that is, it would provide a pressure \cite{Appels1, Diagrams}. 

Finally, we must discuss the range of the $r$ coordinate. The spacetime of the C-metric has different horizons where the function $f(r)$   vanishes, giving rise to several coordinate singularities. This occurs for $r = 2m, \pm 1/A$. The function $\Omega$ also depends on $r$. The points at which it vanishes, i.e. $r = -1/(A\cos\theta)$, define the conformal infinity. We are going to consider $2mA < 1$ for our discussion, but we will see later that the same considerations apply to other values of $2mA$.

We start from $r = 0$ to $r_E = 2m$. We name this the event horizon similar to the Schwarzschild metric. The $r$ coordinate can continue from the event horizon to $r_A = 1/A$, where it reaches the acceleration horizon. This is related to the fact that a uniformly accelerated observer asymptotically reaches the speed of light, and therefore losses causal contact beyond this asymptotic light cone \cite{Appels1}. From here, the conformal infinity is reached for a value of $r$ depending on $\theta$. For $\theta = \pi$, the conformal infinity is just at the acceleration horizon $r = 1/A$ and, for $\pi/2 < \theta < \pi$, it is reached for a positive finite value of $r$. For $\theta = \pi/2$, it is reached at $r = \pm \infty$; but for $\theta < \pi/2$, the conformal infinity  is not reached for any positive value of $r$ and the affine parameter remains finite at $r = \infty$. Actually, the conformal infinity is reached for a negative value of $r \in (-\infty, -1/A]$, being placed at $r = -1/A$ for $\theta = 0$. Then, the region from $r = -1/A$ to $0$ corresponds to a different spacetime, and we can take $r \in [0,\infty) \cup (- \infty, -r_A]$, having only the event and acceleration horizon. The hypersurface $r = -1/A$ is reached at the conformal infinity and only for a single value of $\theta$.

The presence of the acceleration horizon could be eliminated by introducing a negative cosmological constant, which has been discussed, for example, in \cite{Appels1, Appels2} to study the thermodynamics. Indeed, the existence of two horizons originate two different temperatures. In this work, we will focus on the study of the structure of spacetime in the presence of both horizons.

In the following sections we will review the causal structure of the sub-accelerating case, and later we will see what happens in the extremal ($2mA = 1$) and super-accelerating ($2mA > 1$) cases, where we deal with this new singularity at $\theta_0$ with $g (\theta_0) = 0$.

\section{Causal structure of accelerating black holes}
\label{sec-causal}

Once we have presented the C-metric, we will obtain the causal structure of the mentioned accelerating black holes, also distinguishing the case of a positive mass parameter from a negative one. We will therefore see how the causal structure changes with these parameters, comparing with previous studies in the
literature \cite{Diagrams}.

Before starting, we must verify  which sections with  constant  $\theta$ and $\varphi$ are totally geodesic, in other words, whether a geodesic tangent to one of these sections at some point is tangent to it at any other. If we calculate the geodesic equations and fix $\theta$ and $\varphi$ as constants, in the equation for $\theta$ we get:
\begin{equation}
f(r)\dot{t}^2 = {\dot{r}^2}/{f(r)},
\end{equation}
which is the equation for radial null geodesics. In other words, only null geodesics can be purely radial, and hence Penrose diagrams for constant $\theta$ and  $\varphi$ are totally geodesics only for null geodesics. Thus, we can use the null radial geodesics to construct the diagrams, keeping in mind that time-like geodesics will not remain tangent to these  sections.

\subsection{Sub-accelerating black hole: $2mA < 1$ and $m > 0$}

\begin{figure}
\begin{overpic}[width=.32\textwidth]{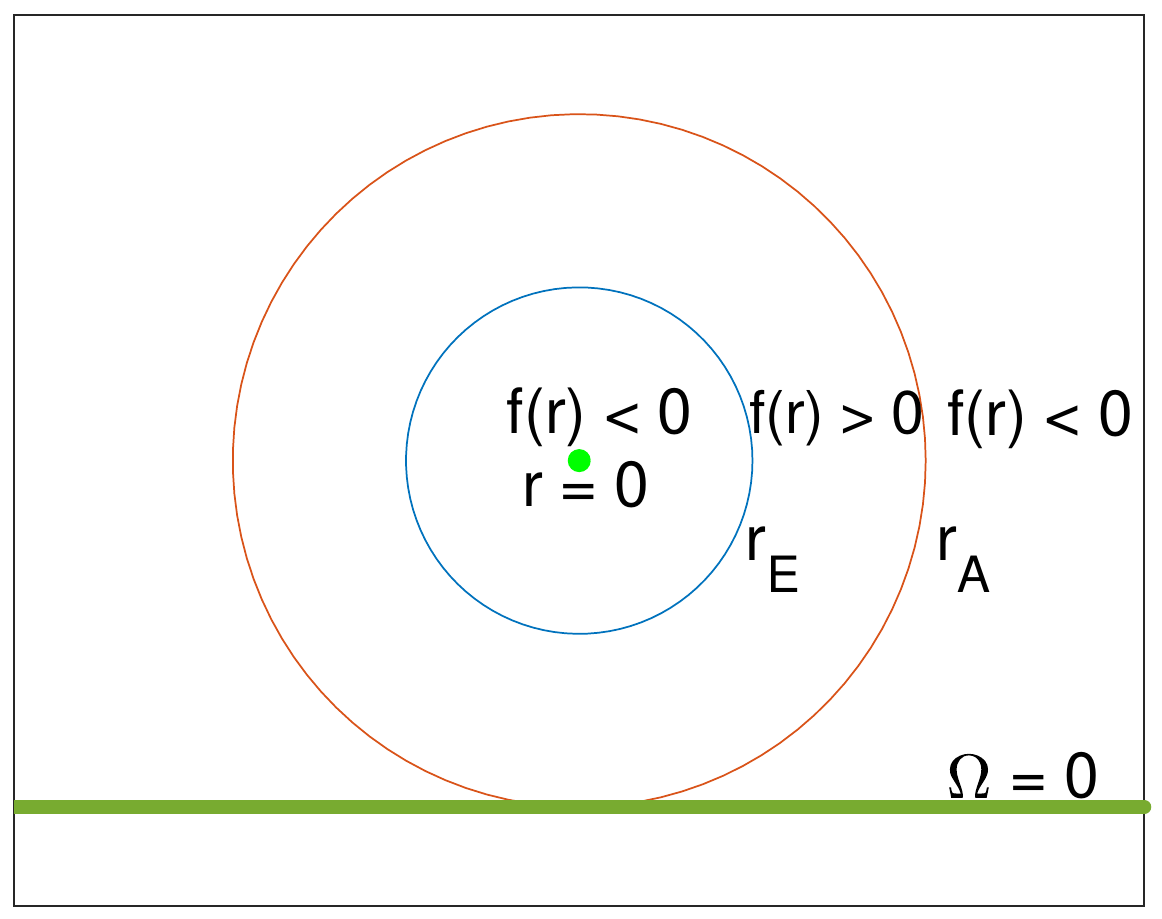}
\put(4,69){\small$2mA < 1$}
\end{overpic}
\begin{overpic}[width=.32\textwidth]{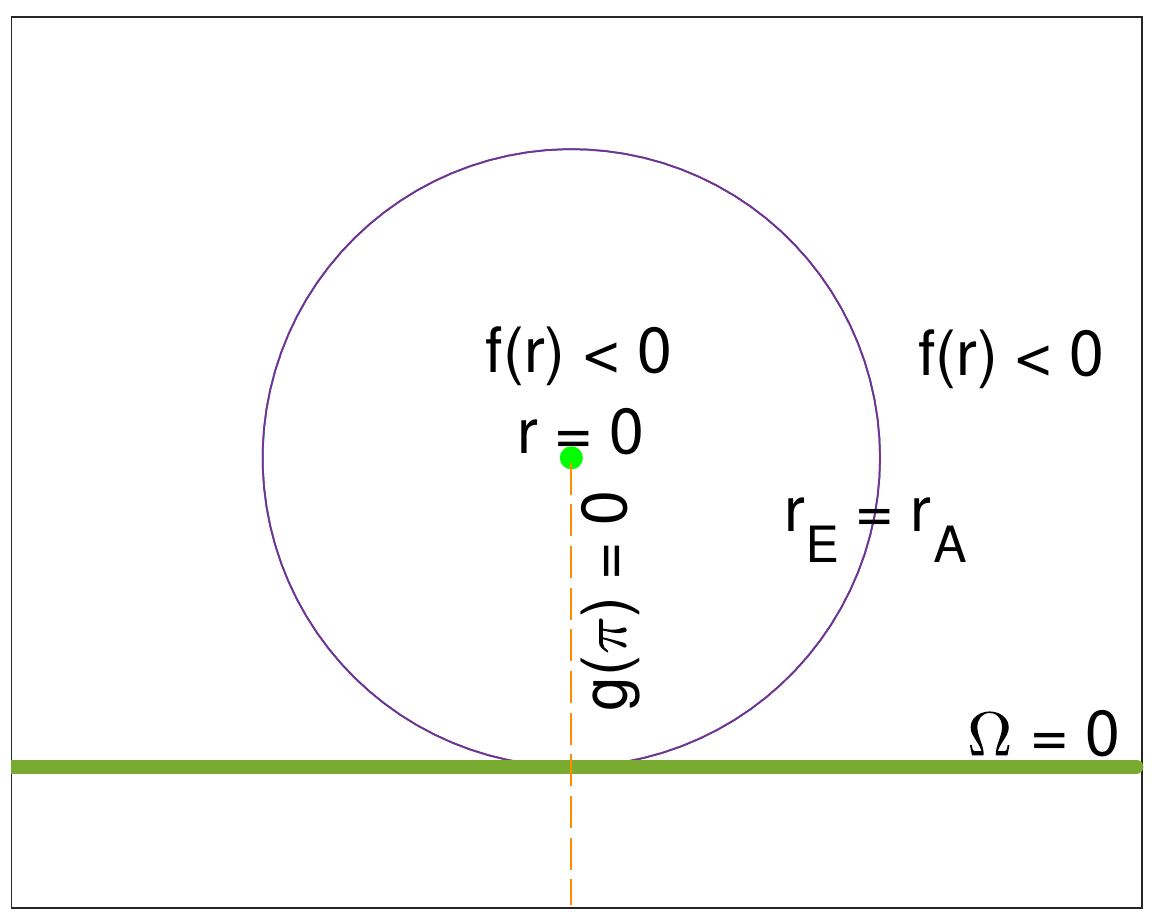}
\put(4,69){\small$2mA = 1$}
\end{overpic}
\begin{overpic}[width=.32\textwidth]{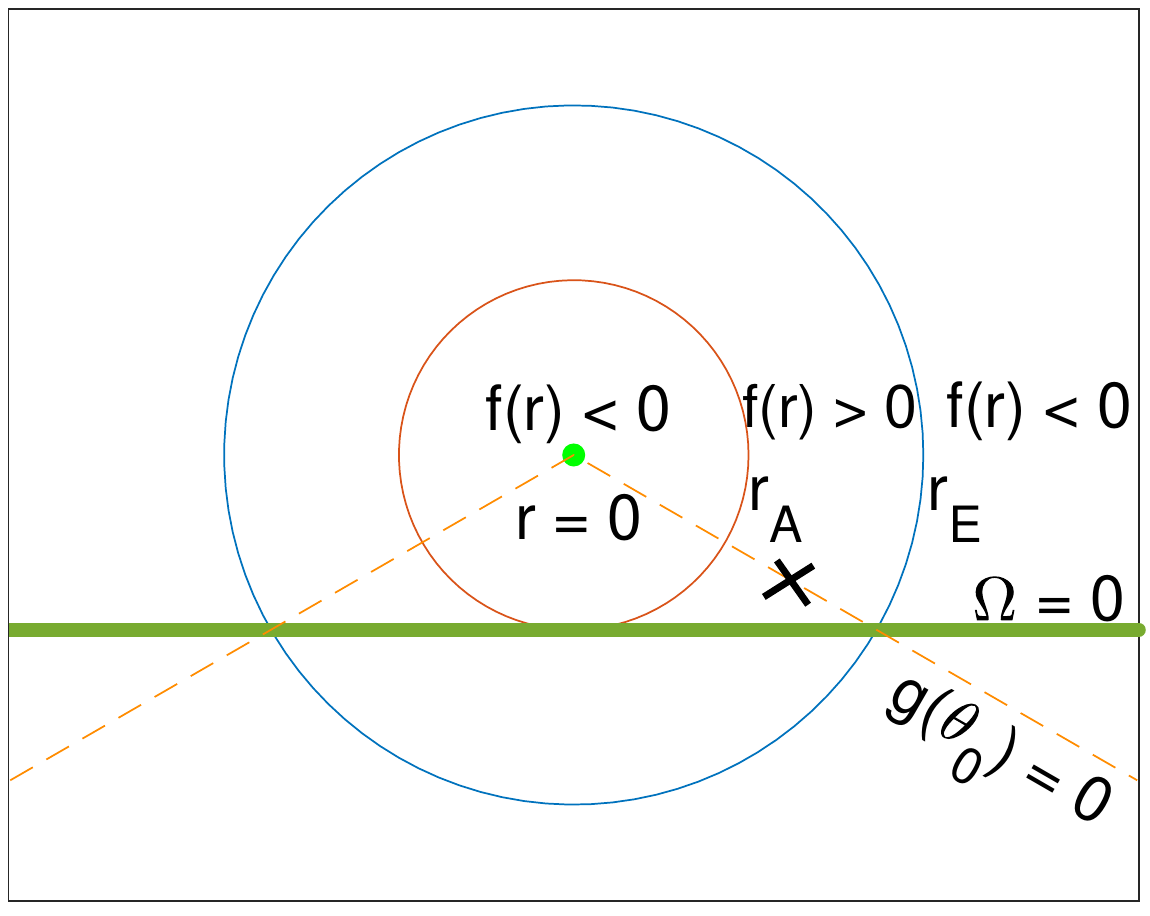}
\put(4,69){\small$2mA > 1$}
\end{overpic}
	\caption{Different structures for accelerating black holes as a function of acceleration.  The vertical and horizontal axes represent $r\cos\theta$  and $r\sin\theta$, respectively. The north semiaxis corresponds to  $\theta = 0$ and the south semiaxis to $\theta = \pi$. The thick (green) line is the conformal infinity reached for finite and positive $r$, the dashed (orange) line gives the values of $\theta$ such that $g(\theta) = 0$, and the solid circles are the event (blue) and acceleration (red) horizons at $r_E$ and $r_A$, respectively. Note that only positive values of $r$ are considered, and recall that in the north hemisphere the conformal infinity is reached for negative values of $r$.}
	\label{F:regions}
\end{figure}

We will briefly review the construction of    Penrose diagrams for the case \mbox{$2mA <1$} and $m>0$, which can be found in \cite{Diagrams}, as it will be helpful when comparing with the other cases later. We start determining the regions and horizons of the spacetime in this case. Since the event horizon is located at $r_E = 2m$ and the acceleration horizon at $r_A = 1 / A$, we have $r_E <r_A$, so the acceleration horizon is further away from the origin than the event horizon. In Figure \ref{F:regions} (left panel), we show the different regions of this spacetime. We have represented $r \cos \theta$ on the vertical axis, while on the horizontal axis we represent $r \sin \theta$. In other words, we are plotting  the section with constant $t$ and $\varphi$. In addition, the thick (green) line corresponds to the conformal infinity where $\Omega = 0$. In the region between horizons, the function $f (r)> 0$, so this region is analogous to the one outside the event horizon within the Schwarzschild geometry \cite{Carroll}. However, for $r < r_E$ or $ r > r_A$, the function $f (r) <0$, so the $r$ coordinate is time-like and the spacetime becomes dynamical. Then, it is impossible to be at rest, just as happens inside the Schwarzschild black hole.

To obtain these Penrose diagrams we must find null coordinates in which  the metric is regular at the horizons, and the conformal infinity is brought to a finite value of the coordinates. To do this, we define the tortoise coordinate $r^*$ such that 
$dr^* = dr/f(r)$, obtaining:
\begin{equation}\label{Tortoise}
r^* = \frac{1}{2\kappa_0}\log|1+Ar| + \frac{1}{2\kappa_A}\log|1-Ar| + \frac{1}{2\kappa_E}\log|r-2m|,
\end{equation}
where
\begin{equation}\label{gravedades}
	\kappa_{E} = {\left(1-4 A^2 m^2\right)}/{(4 m)},\qquad \kappa_{A} = A \left(2 A m - 1\right),
\end{equation}
are the surface gravities  at the event and acceleration horizons, respectively, and
\begin{equation}
\kappa_0 = A (1 + 2 A m).
\end{equation}
We can now define the null  coordinates
\begin{equation}\label{EF}
u = t-r^*, \qquad v = t+r^*,
\end{equation}
in terms of which the metric becomes
\begin{equation}\label{metricanulas}
	ds^2 = \frac{1}{\Omega^2(r,\theta)}\left[-f(r)dudv + r^2d\widetilde{\Omega}^2\right],
\end{equation}
where $r$ is related to $u$ and $v$ through Eqs. \eqref{Tortoise} and \eqref{EF}, and $d\widetilde{\Omega}^2$ is the angular part of the metric. In order to find the maximal analytic extensions of this spacetime, we have to find null coordinates in which the metric is regular at the horizons. This must be done for each horizon separately, so that we will obtain a diagram around each horizon with a region of overlap, which can be merged into one at the end. The diagrams will correspond to sections with constant $\theta$ and $\varphi$. Although they are not totally geodesic as we have discussed, they are valid for null geodesics. In what follows we write only the radial part of the metric.

Let us start from  the region between horizons, and  define there the new pairs of null coordinates
\begin{equation}\label{Kruskal}
	u_i' = -\kappa_i^{-1}e^{-u\kappa_i},\qquad v_i' = \kappa_i^{-1}e^{v\kappa_i},
\end{equation}
where the subscript $i=A,E$ refers to  each horizon. Also, given the expression of $r^*$, each pair of these new null coordinates vanish in their corresponding  horizons. To bring conformal infinity to a finite value of the coordinates, we use the coordinates
\begin{equation}\label{atan}
	\widetilde{u}_i = \arctan(u'_i), \qquad \widetilde{v}_i = \arctan(v'_i).
\end{equation}

\begin{figure}\centering
	\includegraphics[width=.8\textwidth]{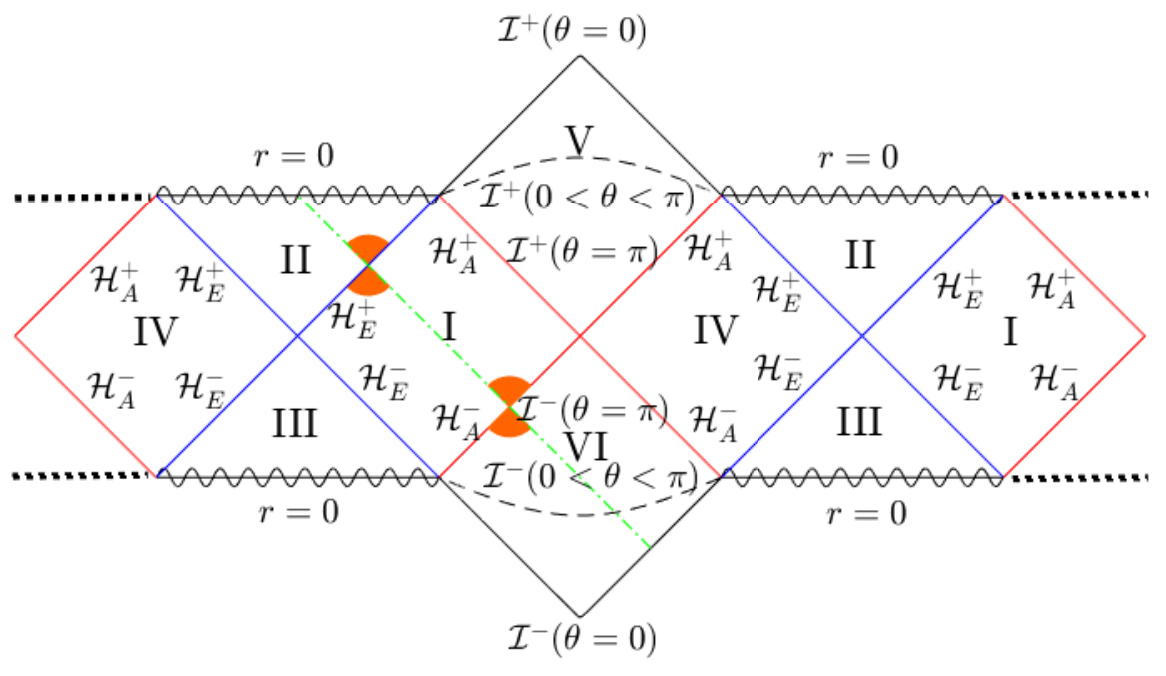}
	\caption{\label{F:Penrose_N} Penrose diagram of the sub-accelerating black hole with $2mA < 1$. $\mathcal{H}_E$ refers to the event horizon, $\mathcal{H}_A$ to the acceleration horizon, and $\mathcal{I}$ to the conformal infinity. The plus superscript means future, while the minus superscript means past. The diagram is actually a superposition of different diagrams for each value of $\theta$. The location of the conformal infinity depends on $\theta$. It is placed at the acceleration horizon for $\theta = \pi$, and closes the diagram with a diamond shape for $\theta = 0$. Since the metric is singular for $\theta = 0, \pi$, the diagrams for those values do not represent the spacetime, so the conformal infinity in those cases has to be understood as a limit when $\theta \rightarrow 0, \pi$. The diagram is invariant under $\varphi$ rotation.}
\end{figure}

Let us deal first with the internal horizon, i.e., in this case the event horizon. In the coordinates $u_E',v_E'$, the radial part  of  the metric becomes
\begin{align}
ds^2  = \frac{-h(r)}{r\Omega^2(r,\theta)}du'_Edv'_E,\qquad
h(r) =  (1-A r)^{  [6+(A m)^{-1}]/4} (1+A r)^{ [6-(A m)^{-1}]/4},
\end{align}
so that it is perfectly regular at the event horizon $r_E = 2m$ (but not at $r_A = 1/A$). Outside this horizon we have $u'_E < 0$ and $v_E ' > 0$. On the other hand, the radial null geodesics parametrised by $ 
u_E'$ or $v_E'$ reach the horizon at finite affine parameter. Therefore, we can extend  $u_E'$ and $v_E'$ to $(-\infty,+\infty)$, so that we will have four regions. These regions are associated with regions I, II, III, and IV on the Penrose diagram of Figure \ref {F:Penrose_N}.
The coordinates $r$ and $t$ in each region are related with the coordinates $u'_E$ and $v'_E$ by
\begin{align}\label{rE}
	|u'_Ev'_E|& = (\kappa_E)^{-2}e^{2\kappa_E r^*(r)}, \qquad t = (2\kappa_E)^{-1}\log\left| {v_E'}/{u_E'}\right|.
\end{align}
We can see in the diagram the curves corresponding to constant $r$. In this figure, the past and future event horizons $\mathcal{H}_E^\pm$ correspond to the axes $u_E'=0$ and $v_E'=0$.

The extension through the acceleration horizon  $\mathcal{H}_A^\pm$, can be carried out in an  entirely analogous way. The coordinates $r$ and $t$ and  the corresponding null coordinates $u'_A$ and $v'_A$ are now
\begin{align}\label{rA}
	|u'_Av'_A| &= (\kappa_A)^{-2}e^{2\kappa_A r^*(r)}, \qquad t = (2k_A)^{-1}\log\left| {u_A'}/{v_A'}\right|,
\end{align}
valid in regions I, IV, V and VI of the Penrose diagram  of  Figure \ref{F:Penrose_N}. The two sets of coordinates overlap in region I  and region IV. Then, the resulting Penrose diagram is the result of infinitely concatenating the two extensions described above via these overlaps.

Figure \ref{F:Penrose_N} actually shows a superposition of diagrams for all angles $\theta$. For each value of $\theta$, the conformal infinity $\mathcal{I}^\pm$ is reached at $r = -(A \cos \theta)^{- 1}$, so its location on the diagram depends on $\theta$. For $\theta = \pi$, it is reached just at the acceleration horizon. So there would be no regions V and VI. For $0 < \theta < \pi$, it is reached for a value of $r \in (r_A,\infty) \cup (- \infty, -r_A)$ as we discussed above. In this case, conformal infinity is represented by a dashed curve that closes the regions V and VI before forming the diamond structure. Finally, for $\theta = 0$, conformal infinity is placed at $r = -r_A$, closing the diagram with a diamond shape. We must take into account that the spacetime is singular in the axes $\theta = 0, \pi$ due to the conical singularity (although one but not both can be regularized), so the Penrose diagrams in these cases have to be understood as a limit when $\theta \rightarrow 0, \pi$.

In the diagram of Figure \ref{F:Penrose_N}, we have represented a null geodesic with a dash-dotted (green) line that starts from conformal infinity in the past, crosses the past acceleration horizon and the future event horizon, and ends at the singularity $r = 0$. In region VI, $f(r)<0$ and any observer would be forced to move towards the acceleration horizon and cross it. In region I, $f(r) > 0$, and therefore the coordinate $r$ is spacelike, and an observer could remain between both horizons in a similar way to the region outside the Schwarzschild black hole. In region II, $f(r)<0$ and every observer is   forced to move towards the singularity at $r = 0$. However, we must take this discussion with care, since we have discussed that timelike observers are not totally geodesic for constant $\theta$ and $ \varphi$, and therefore their movement could go out of the diagram towards other sections of constant angles. In \cite{Diagrams}, 3D Penrose diagrams that allow to study the dependence with the angle $\theta$ can be found. We also have a white hole in III, from which any observer exits through the past horizon. In region V the null geodesics exit through the acceleration horizon to necessarily reach   conformal infinity. Region IV is similar to region I.

Many of these areas, such as the white hole or past conformal infinity, appear because we are not considering the dynamical formation of the black hole. To better understand the physics of these objects, it will be necessary to study in future works the causal structure of accelerating black holes formed by gravitational collapse of stars. 

\subsection{Super-accelerating black hole: $2mA > 1$ and $m > 0$}

The super-accelerating black hole is obtained when we choose an acceleration value such $2mA > 1$. The horizons and regions of the spacetime are represented in the Figure \ref{F:regions} (right panel). In this case, the acceleration horizon is inside the event horizon. However, again the function $f(r) > 0$ between both horizons, and $f(r) < 0$ on the rest of the spacetime. In this case a singularity associated with $g(\theta) = 0$ appears. This is reached at a value $\theta_0\in (\pi/2,\pi)$ of the polar angle which depends on the value of $2mA$. This gives rise to a possibly singular area represented by a dashed (orange) line which, having revolution symmetry, determines a cone.

In order to analyse the nature of this singularity, we make the change $x = \cos\theta$, so the metric reads:
\begin{align}
	ds^2 &= \frac{1}{\Omega^2(r,\theta)}\left[-f(r)dt^2 + \frac{dr^2}{f(r)} + r^2d\widetilde{\Omega}^2\right],\qquad
	d\widetilde{\Omega}^2  = \frac{dx^2}{G(x)}+G(x)d\varphi^2,
\end{align}
where $G(x) (1 - x^2)(1 + 2mAx)$. This function $G(x)$ has roots at $x =\pm 1$, and another one in $x = -1/2mA$, which is inherited from the function $g(\theta)$ in \eqref{eq-f} and causes the singularity we are dealing with. We call this last root $x_0 = \cos \theta_0$. We expand the metric around the singularity $x = x_0$, at a value of the radial coordinate $r = r_0$ such that $f(r_0) > 0$ (this point is marked in the right panel of Figure \ref{F:regions} with a cross). We shall expand on each side of the singularity, so that $x = x_0(1\pm\zeta^2)$, with $0 \leq \zeta \ll 1$. The positive sign is taken to expand in the region $\theta <\theta_0$, and the negative sign for the region $\theta > \theta_0$. Then, $G[x_0(1\pm\zeta^2)] \simeq \pm C \zeta^2$, since $G(x_0) = 0$, with $C$ being a positive constant. On the other hand, in terms of the new coordinate $\zeta$ the angular metric $d\widetilde{\Omega}^2$ reads:
\begin{equation}
	d\widetilde{\Omega}^2 = \pm \left(K_1 d\zeta^2 + K_2\zeta^2 d\varphi^2\right),
\end{equation}
where $K_1$ and $K_2$ are two positive constants depending on $C$, whose particular values are irrelevant for our discussion.

We also expand the $r$ coordinate around the value $r_0$ so that $r = r_0 + \rho$. The function $f(r) \simeq f_0 = \text {constant} > 0$ at lowest order in $\rho$, so it can be reabsorbed in the coordinates $t$ and $\rho$. On the other hand, the factor $\Omega^2(r_0, \theta_0) = \text {constant} \neq 0$ can also be absorbed in $t$ and $\rho$. For the $r^2$ factor that multiplies $d\widetilde{\Omega}^2$, we take it as $r_0^2$. So, this constant $r_0^2$, the constant $\Omega$ factor, and the positive constants $K_1$ and $K_2$ can be reabsorbed in $\zeta$ and $\varphi$ respectively, leaving the metric as:
\begin{equation}
	ds^2 = -dt^2 + d\rho^2 \pm d\zeta^2 \pm \zeta^2\varphi^2.
\end{equation}

Finally, we can analyse the null geodesics near the singularity by taking $\rho$ and $\varphi$ constants. The metric for $\theta < \theta_0$, is simply the Minkowski metric in two dimensions, $ds^2 = -dt^2 + d\zeta^2$. Then, the null geodesics reach the singularity at $\zeta = 0$ with a finite affine parameter. However, it is not possible to continue the geodesics due to the change of signature in the angular coordinates, so this singularity is an essential singularity. Then, for the super-accelerating case we have $\theta \in [0,\theta_0)$. Note that although the conformal infinity can be placed inside the event horizon, that would correspond to $\theta > \theta_0$, so again the conformal infinity is only reached outside the outer horizon, and the discussion for the range of $r$ is equivalent to that  in the case $2mA < 1$.

\begin{figure}\centering
	\includegraphics[width=.8\textwidth]{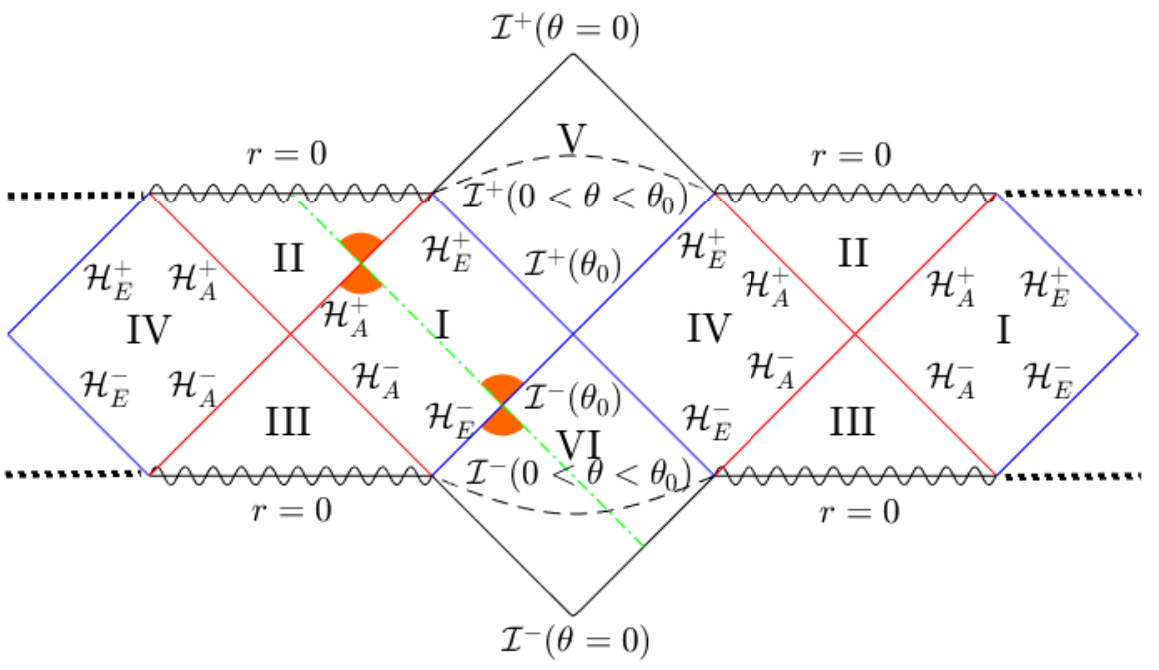}
	\caption{\label{F:Penrose_R} Penrose diagram of the super-accelerating black hole with $2mA > 1$. $\mathcal{H}_E$ refers to the event horizon, $\mathcal{H}_A$ to the acceleration horizon, and $\mathcal{I}$ to the conformal infinity. The diagram is actually a superposition of different diagrams for each value of $\theta$. The location of the conformal infinity depends on $\theta$. It is placed at the event horizon for $\theta = \theta_0$, and closes the diagram with a diamond shape for $\theta = 0$. Since the metric is singular for $\theta = 0, \theta_0$, the diagrams for those values do not represent the spacetime, so the conformal infinity in those cases has to be understood as a limit when $\theta \rightarrow 0, \theta_0$. The diagram is invariant under $\varphi$ rotation.}
\end{figure}

Once we have analysed the regions and singularities for the super-accelerating case, we can determine the causal structure, with the help of the calculations made  for the sub-accelerating black hole. In this case we obtain the same tortoise coordinate \eqref{Tortoise}. We use the same definition of the Kruskal-Szeckeres-type coordinates than in Equation \eqref{Kruskal} to regularize the horizons. Beginning again in the area between horizons, in this case $r_A < r < r_E$, extending the coordinates to the entire real line and performing a conformal compactification according to the changes of coordinates given by Equation \eqref{atan}, we obtain the diagrams for each horizon. Concatenating them, we obtain the Penrose diagram of Figure \ref{F:Penrose_R}.

To represent the curves of constant $r$, we define this coordinate from $u'$ and $v'$ for the regions surrounding each horizon as in Equations \eqref{rE} and \eqref{rA}. The diagram is similar to the one in the sub-accelerating case with $2mA < 1$. Again, we show a superposition of different diagrams for each value of $\theta$. The location of the conformal infinity depends on~$\theta$. However, there are two major differences. The first one is that the event horizon $\mathcal{H}_E^\pm$ is interchanged with the acceleration horizon $\mathcal{H}_A^\pm$. The second one is that whereas in the previous case the conformal infinity was reached at the outer horizon for $\theta = \pi$, now it is reached at $\theta_0$. This is the maximum value of $\theta$ with $\cos\theta_0 = -1/2mA$, which makes the metric singular. Then the conformal infinity is placed at the event horizon as it can be seen in Figure \ref{F:regions} (right panel). Both horizons play the role of point of no return as it is shown by the dash-dotted (green) null geodesic. Then, although they are interchanged with respect to the sub-accelerating case, the discussion for the causal structure in each of the regions is the same. We obtain again that for $\theta = 0$ conformal infinity is reached at $r = -r_A$. Again the metric is singular at the two extreme values of $\theta$ ($\theta = 0,\theta_0$), so their diagrams do not represent the spacetime, but are limiting cases.

\subsection{Extremal accelerating black hole: $2mA = 1$ and $m > 0$}

If we take $2mA = 1$, the event and acceleration horizons are located at the same position, becoming an extremal black hole. This is shown in Figure \ref{F:regions} (central panel). The function $f(r) < 0$ both inside and outside the horizon, so an observer could never be at rest. In this case, for the axis $\theta = \pi$, the singularity associated with $g(\theta) = 0$ appears. It is represented with a dashed (orange) line.

We fix the acceleration as $A = 1 / (2m)$, so both horizons merge at $r_E = r_A = 2m$. The tortoise coordinate for this case can then be defined as

\begin{equation}\label{Tortoise_E}
	r^* = \frac{2m^2}{r-2m} - \frac{m}{2}\log|r-2m| + \frac{m}{2}\log|r+2m|.
\end{equation}

As the function $f(r) < 0$ in any region of spacetime of this metric, the $r$ coordinate is globally timelike, and in the same way so is the coordinate $r^*$. In order to construct null coordinates oriented towards the future, we must take into account the orientation of the tortoise coordinate, which depends on whether we describe a black hole or a white hole.

In the case of a black hole, the singularity $r = 0$ is placed in the future, so the $r$ coordinate is oriented towards the past. Over the entire range of the $r$ coordinate we are considering, that is $r \in [0, \infty) \cup (- \infty, -r_A]$, the $r^*$ coordinate is decreasing with $r$ and, therefore, it is oriented to the future. Thus, if we want null coordinates oriented towards the future, we can take:
\begin{equation}
	u = r^*-t, \qquad v = r^* + t.
\end{equation}

The metric in these coordinates takes the form of Equation \eqref{metricanulas} except for a sign in the radial part. Then, these coordinates allow us to construct the Penrose diagram, after performing the conformal compactification according to the coordinates $\widetilde{u}$ and $\widetilde{v}$ defined by:
\begin{equation}
	\widetilde{u}= \arctan(u), \qquad \widetilde{v}= \arctan(v).
\end{equation}

In these coordinates the radial part of the metric reads:
\begin{equation}
    ds^2 = \frac{h}{\Omega^2(r,\theta)} d\widetilde{u}d\widetilde{v}, \qquad 
    h = \frac{(1-r^2/(4m^2))(1-2m/r)}{\cos^2(\widetilde{u})\cos^2(\widetilde{v})},
\end{equation}
where $h$ is a negative function that never vanishes. Indeed, the numerator only vanishes at $r=2m$ and so does the denominator, so that its limit when $r\to 2m$ remains finite. 

\begin{figure}\centering
	\includegraphics[width=.8\textwidth]{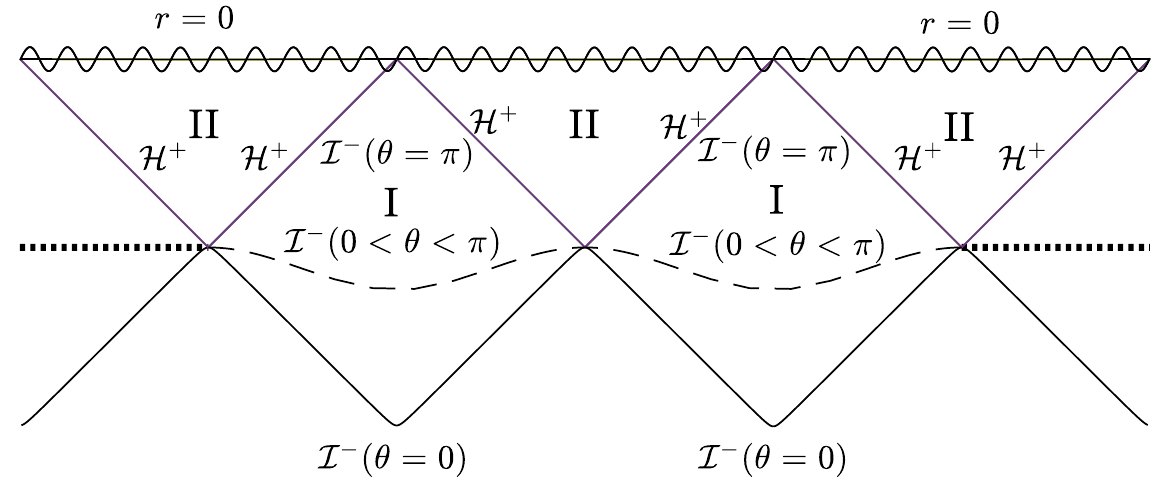}
	\caption{\label{F:Penrose_C_E} Penrose diagram of the extremal accelerating black hole ($2mA = 1$). In this case the singularity $r = 0$ is only in the future. $\mathcal{H}^+$ refers to the future horizon, and $\mathcal{I}^-$ to the past conformal infinity. The diagram is actually a superposition of different diagrams for each value of~$\theta$. The location of the conformal infinity depends on $\theta$. It is placed at the horizon for $\theta = \pi$, and closes the diagram with a diamond shape for $\theta = 0$. Since the metric is singular for $\theta = 0, \pi$, the diagrams for those values do not represent the spacetime, so the conformal infinity in those cases has to be understood as a limit when $\theta \rightarrow 0, \pi$. The diagram is invariant under $\varphi$ rotations.}
\end{figure}

The curves   $r = $ constant are obtained from $u + v = 2r^*(r)$. We represent the Penrose diagram in Figure \ref{F:Penrose_C_E}. In this case, given the periodicity of the tangent function, we obtain an infinite series of zones in the diagram, although only of two different types, named as I and II. Region II corresponds to the interior of the black hole, while region I is the exterior, in which the past conformal infinity is located. As in previous cases, we show a superposition of different  diagrams for different constant values of $\theta$. In these diagrams, the conformal infinity depends on the chosen value of $\theta$.

Since there is only one horizon $\mathcal{H}$, and on each side the coordinate $r$ is timelike because $f(r)< 0$, every observer is forced to enter the black hole, and continue moving towards the singularity. Thus, in this case it is not possible for an observer to move away from the black hole. Regarding the past of the spacetime, it should be noted that we are not considering the dynamical formation of the black hole, so it should be taken with care as in the non-extremal cases.

As in the previous cases, this metric could also represent a white hole. To get the diagram in that case, the coordinate $r$ should be oriented towards the future, leaving the singularity $r = 0$ in the past. Thus, $-r ^*$ would be the future oriented coordinate and we would define the null coordinates as $u = -r^* - t$ and $v = -r^* + t$. This is nothing more than a temporal reflection of the discussed black hole in Figure \ref{F:Penrose_C_E}.

\subsection{Accelerating black hole with $m<0$}

In the Schwarzschild black hole a negative mass parameter makes the event horizon disappear, so that the curvature singularity is naked, affecting the predictability of spacetime \cite{SchwM2}. In this section we would like to discuss how accelerating the black hole protects from this singularity.

\begin{figure}\centering
	\includegraphics[width=.4\textwidth]{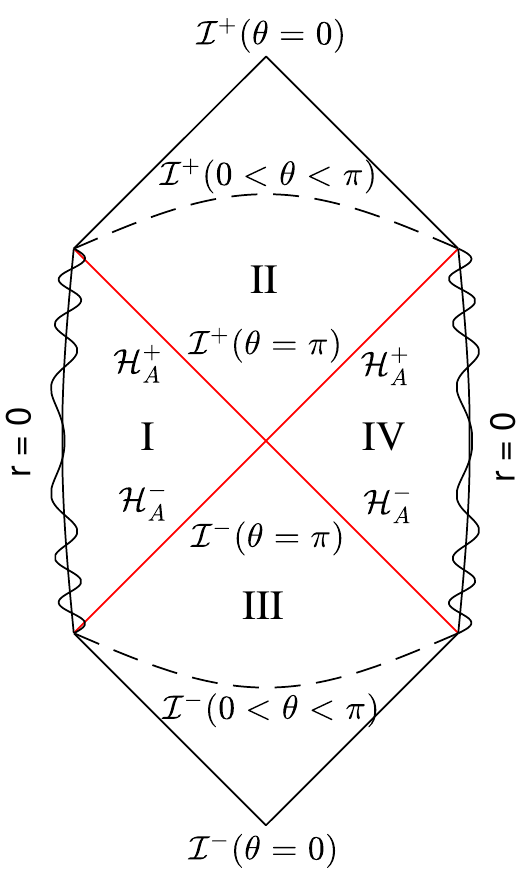}
	\caption{\label{F:Penrose_M} Penrose diagram of the accelerating black hole with $|2mA| < 1$ and $m < 0$. $\mathcal{H}_A$ refers to the acceleration horizon, and $\mathcal{I}$ to the conformal infinity. The plus superscript means future, while the minus superscript means past. As in previous figures, we show a superposition of different $\theta = \text{constant}$ diagrams, whose conformal infinity depends on the chosen value of $\theta$. The diagram is invariant under $\varphi$ rotation.}
\end{figure}

Let assume $|2mA| < 1$ for simplicity, so that the singularity at $g(\theta) = 0$ does not appear. Having $m < 0$ the event horizon could be at $r = -2|m|$, but as this is further from the conformal infinity, this horizon disappears. Then, we only have the acceleration horizon. To obtain the Penrose diagram, it is simply necessary to change $m$ to $-|m|$ in the calculus for the acceleration horizon with $2mA < 1$. Thus, we obtain the diagram in Figure \ref{F:Penrose_M}. The change in the mass parameter only affects the structure inside the acceleration horizon, so we can conclude that the causal structure beyond this horizon is the same as in the positive mass case shown in Figure \ref{F:Penrose_N}.

We can observe that in this case the singularity $r = 0$ is timelike, so an observer in region I or IV inside the horizon is not forced to go towards it (the coordinate $r$ is spacelike). This singularity is unprotected in regions I and IV as would happen in the Schwarzschild black hole. However, regions II and III are protected by the acceleration horizon. So, for $r > r_A$, the curvature singularity does not affect to the predictability of the spacetime, although the conical singularity at the axis does.

The causal structure in this case is similar to the one obtained for two objects accelerating in opposite directions in flat spacetime \cite{Diagrams}, with the exception that there is a singularity at the origin of each of these two objects, located at $r = 0$. This reminds the case of the Schwarzschild black hole with negative mass, whose Penrose diagram is like Minkowski with a singularity at the origin \cite{SchwM1}.

Had we taken $|2mA| \geq 1$, a similar diagram would be obtained. However, in this case  there appears a singularity at $\theta_0$ as happened in the super-accelerating positive-mass case. Therefore $\theta $ must be restricted to the interval $\theta \in (\theta_0, \pi]$. The hypersurface $r = -2|m|$, which would give rise to another horizon, lies at the conformal infinity for $\theta = \theta_0$ and beyond it for all other values of $\theta$. Then, we would have only an acceleration horizon again. 

In this work, for positive mass $m$, we have considered the radial coordinate $r \in [0,+\infty) \cup (- \infty, -r_A]$, but if we did the discussion for the range of $r$ starting from $r = 0^-$ towards negative values until we reach conformal infinity, then we would describe the spacetime at the other side of the conformal infinity. If we then do the change of coordinates $r \rightarrow -r$, $\theta \rightarrow \pi-\theta$ and consider the new $r$ coordinate from $0^+$ towards positive values until reach conformal infinity, we obtain the C-metric with $-m$ instead of~$m$. Thus, the spacetime of the accelerating black hole with negative mass is equivalent to the spacetime that the C-metric would describe with $m >0$ in the region that goes from conformal infinity up to $r = 0^-$.

\section{Conclusions}
\label{sec-conlusions}

We have studied the properties of the C-metric, which describes the spacetime of accelerating black holes, in order to focus in its causal structure. This metric has a conical singularity associated with a cosmic string responsible for the acceleration. Also, this metric has both an event horizon related to a black hole, and an acceleration horizon.

While in the literature the value of the acceleration is usually restricted such that \mbox{$2mA<1$}, we have allowed the acceleration to take any value in our work. Thus, we have studied the singularity that appears at a certain value of the angular coordinate $ \theta = \theta_0 $ when $ 2mA \geq 1 $, observing that it is an essential singularity of the metric.

We have obtained that purely radial geodesics can only be null, and from them we have constructed the Penrose diagrams for accelerating black holes with any value of the acceleration. In the super-accelerating case ($2mA > 1$), the acceleration and event horizons exchange their roles. In addition, an unprotected singularity appears for $ 2mA> 1 $ at an angle $ \theta_0\in(\pi/2,\pi) $. On the other hand, for the extremal case with $2mA = 1$, we have obtained independent Penrose diagrams describing either a black hole or a white hole, where an observer can never be static and either enters in a black hole or go out from a white hole.

We have also studied the accelerating black hole with $ m <0 $, observing that the acceleration protects the curvature singularity at $r = 0$ unlike the Schwarzschild case. We have observed that the spacetime is like Minkowski's for two objects accelerating in opposite directions, but with a singularity at the origin of each of them. 
Furthermore, we have also shown that  the positive-mass C-metric  on the other side of conformal infinity is equivalent to the accelerating black hole with negative mass.

\acknowledgments
This work was partially supported by the MICINN (Spain) projects PID2019-107394GB-I00 and PID2020-118159GB-C44. S.A.O. thanks IPARCOS-UCM and the CAM/FEDER Project No.S2018/TCS-4342 (QUITEMAD-CM) for funding support.
 
\bibliography{MiBiblio}
\bibliographystyle{unsrt}

\end{document}